\documentclass[twocolumn,showpacs,amsmath,amssymb,pra]{revtex4}

\usepackage{bbm}

\usepackage{graphicx}
\usepackage{dcolumn}
\usepackage{amsmath}
\usepackage{bm}
\usepackage{color}
\usepackage{mathrsfs}
\usepackage{epstopdf}
\usepackage{amssymb}
\usepackage{subfigure}
\usepackage{enumitem}
\usepackage{multirow}
\usepackage{exscale}
\usepackage{relsize}
\usepackage{dsfont}
\usepackage{pst-node}
\usepackage{tikz-cd}

\usepackage{float}
\usepackage{mathrsfs}

\newcommand{\be}{\begin{equation}}
\newcommand{\ee}{\end{equation}}
\newcommand{\bey}{\begin{eqnarray}}
\newcommand{\eey}{\end{eqnarray}}
\newcommand{\bw}{\begin{widetext}}
\newcommand{\ew}{\end{widetext}}

\newcommand{\ba}{\begin{array}}
\newcommand{\ea}{\end{array}}
\newcommand{\bi}{\begin{itemize}}
\newcommand{\ei}{\end{itemize}}
\newcommand{\bem}{\begin{enumerate}}
\newcommand{\eem}{\end{enumerate}}

\begin{document}

\title{Many-body systems with $\mathrm{SU}(1,1)$ dynamical symmetry: from dynamics to thermodynamics based on the trace formula}

\author{Zhaoyu Fei} \email[Email: ]{zyfei@gscaep.ac.cn}
\affiliation{Graduate School of China Academy of Engineering Physics,
No. 10 Xibeiwang East Road, Haidian District, Beijing, 100193, China}

\author{C. P. Sun} \email[Email: ]{suncp@gscaep.ac.cn}
\affiliation{Graduate School of China Academy of Engineering Physics,
No. 10 Xibeiwang East Road, Haidian District, Beijing, 100193, China}
\affiliation{Beijing Computational Science Research Center, Beijing, 100193, China}

 \date{\today}

\begin{abstract}
For a quantum (many-body) system with dynamical symmetry described by a given Lie group, we study the trace of exponential operators with complex coefficients in one of the irreducible subspaces in terms of the boson realization. By using this approach, for compact groups, we obtain the result of the trace that is consistent with the well-known Weyl character formula. For non-compact groups (with $\mathrm{SU}(1,1)$ group as an application), convergent condition of the trace is also obtained. This approach may be a powerful tool to study the thermodynamics of quantum (many-body) systems in equilibrium states or nonequilibrium processes.
\end{abstract}

\maketitle

\section{Introduction}

For quantum (many-body) systems with dynamical groups, both the dynamics (time-evolution operator) and the thermodynamics (partition function) can be approached based on the trace of exponential operator (trace formula). Extending the domain of the partition function to the complex plane, the trace formula corresponds to the Lee-Yang zeros (grand-canonical ensemble)~\cite{yang1952,lee1952} or Fisher zeros (canonical ensemble)~\cite{fisher1965} which establish a connection between the properties (the zeros) of a partition function for a finite size system and phase transitions that may occur in the thermodynamic limit.  Moreover, such a trace formula is also involved in the study of quantum (many-body) systems in nonequilibrium processes, such as the work statistics~\cite{aq2000,ja2000}, full-counting statistics in electric circuit~\cite{kinder2003}, Loschmidt echoes~\cite{goussev2012}, and dynamical phase transition. It is worth mentioning that both the Lee-Yang zeros and the characteristic function of work have been observed in experiments  by measuring quantum coherence of a probe spin~\cite{peng2015} and the qubit-assisted Ramsey-like scheme~\cite{mazzola2013} respectively.

Since the spectra and evolution of the system are solvable, it is reasonable to expect that the trace formula is also solvable. Actually, the Levitov's formula~\cite{levitov1996,klick2003} and the trace formula for quadratic Hamiltonian~\cite{fei2019} have already been discovered for identical-particle systems. However, these formulas, which are valid when the Hilbert space consists of Fock states, can not be applied when the trace formula in an irreducible subspace is concerned with.
%
For compact Lie groups, the irreducible representation is finite-dimensional. Thus, the trace formula is equivalent to the characters of Lie groups. By using the Weyl character formula~\cite{weyl1953,fulton2004}, there is a systematical approach to calculating the characters~\cite{fall2015}. For non-compact Lie groups, thing are quite different due to the infinite dimensions of the Hilbert space. The trace formula is not conjugate-invariant or  not convergent in some cases. Then, the trace formula is not equivalent to the (global) characters of Lie groups~\cite{knapp2016,kirillov2012} as the latter are always  conjugate-invariant and convergent.

In this paper, we want to overcome this conundrum by using the boson realization of Lie groups. The boson realization starts from  Schwinger's representation of $\mathrm{SU}(2)$ and is  further applied to other Lie groups such as $\mathrm{SU}(1,1)$~\cite{gerry1991} and $\mathrm{SU}(N)$~\cite{mathur2010}. In addition, with the help of the projection operator  and the coherent-representation technology, we transform the trace formula into a gaussian integral in a complex plane whose result and convergent condition are easy to obtain. 
As an application, we obtain the result and the convergent condition of the trace formula for the non-compact group: $\mathrm{SU}(1,1)$. This group is of importance in the study of quantum many-body systems since it is a broad family of interacting many-body systems that are analytically solvable~\cite{beau2020}.

\section{Preliminaries}

%


We study a boson realization of group $G$ means that the corresponding Lie algebra $\mathfrak{g}$ can be realized by the quadratic form of bosonic operators. 
Let $a_i,a^\dag_j,i,j=1,\cdots,r$ denote the number of $r$ bosonic annihilation and creation operators, i.e., $[a_i,a_j^\dag]=\delta_{ij}$. The associated Hilbert space $\mathcal{H}$ is the Fock space $\{|n_1,\cdots,n_r\rangle\}\equiv\{|\bm{n}\rangle\}$. In the boson realization of group $G$, for the elements in the Lie algebra ($\forall ix\in \mathfrak{g}$), there is a $2r\times 2r$ symmetric matrix $M(x)$ such that
\be
\label{e1}
\hat x=\bm{\alpha}M(x)\bm{\alpha}^T
\ee
is a representation of $x$ in the Hilbert space, where $T$ denotes the matrix transpose, $\bm{\alpha}=(\bm{a},\bm{a}^\dag)$ and $\bm{a}=(a_1,\cdots,a_r)$, $\bm{a}^\dag=(a_1^\dag,\cdots,a_r^\dag)$ are the vectors of annihilation and creation operators. In physics, Eq.~(\ref{e1}) is the second quantization of one-body operators. We also set
\be
\omega=\left(\begin{matrix}
   0 & 1_r  \\
   1_r & 0
 \end{matrix}\right),
\ee
with the $r\times r$ identity matrix $1_r$
and require that $\omega M(x)$ is hermitian, which preserves the unitarity of this representation. In addition, according to Refs.~\cite{balian1969}, there is a symplectic matrix $2\tau M(x)$ (called the symplectic representation of $x$) associated with $M(x)$, where
\be
\tau=\left(\begin{matrix}
   0 & 1_r  \\
   -1_r & 0
 \end{matrix}\right).
\ee

Since the Hilbert space is the representation space of group $G$, it can be decomposed as the direct sum of the irreducible representation of $G$, i.e., $\mathcal{H}=\bigoplus_\mu\mathcal{H}_\mu$. Here, $\mathcal{H}_\mu=\{|\mu,h\rangle\}$ is one of the irreducible subspaces, $\mu$ indicates different irreducible representations and $h$ denotes the eigenvalues of the Cartan subalgebra of group $G$ (denoted by $\mathfrak{h}$), i.e., $\hat y|\bm\mu,\bm h\rangle=y(\bm h)|\bm\mu,\bm h\rangle$ for $\hat{y}\in\mathfrak{h}$ and some function $y(\bm h)$.


Due to the linearity of Lie algebra, the Hilbert space $\mathcal{H}$ can be also regarded as the representation space of the complexification of the Lie algebra $\mathfrak{g}$ (denoted by $\mathfrak{g}_\mathbb{C}$). If the corresponding complex group $G_\mathbb{C}$ is a complex connected Lie group, $G_\mathbb{C}$ is weakly exponential, namely $\bigcup_{x\in\mathfrak{g}_\mathbb{C}}\exp(x)$ is a dense set in $G_\mathbb{C}$~\cite{hofmann1978}. In this case, the Hilbert space $\mathcal{H}$ is also regarded as the representation space of $G_\mathbb{C}$. In this paper, we aim to calculate the trace formula of $G_\mathbb{C}$ in the subspace $\mathcal{H}_{\bm \mu}$, namely to obtain the expression of
\be
\label{e4}
\mathrm{Tr}_{\bm\mu}[\exp(\hat x)]=\sum_{\bm h}\langle\bm\mu,\bm h |\exp(\hat x)|\bm\mu,\bm h\rangle
\ee
and the convergent condition of Eq.~(\ref{e4}). When $\hat x$ is skew-Hermitian, $\exp(\hat x)$ denotes the time-evolution operator of a quantum system. When $\hat x$ is Hermitian, Eq.~(\ref{e4}) is the partition function in the irreducible subspace. In general, $\hat x$ is a combination of Hermitian operator and skew-Hermitian operator. Then, Eq.~(\ref{e4}) corresponds to the Lee-Yang zeros~\cite{yang1952,lee1952} or Fisher zeros~\cite{fisher1965}. Moreover, since the representation matrices preserve the group multiplication, $\exp(\hat x)$ is also able to denote the product of several exponential operators. Thus, Eq.~(\ref{e4}) corresponds to the generator of correlation functions, or the distribution for some statistics of dynamical systems, such as the work statistics~\cite{aq2000,ja2000}, full-counting statistics in electric circuit~\cite{kinder2003}.

\section{projection onto the irreducible subspace}

Because the trace in the irreducible subspace is hard to deal with, we introduce the projection operator $\hat {P}_{\bm \mu}$
\be
\hat {P}_{\bm \mu}|\bm \mu',\bm h'\rangle=\delta_{\bm \mu \bm \mu'}|\bm \mu,\bm h'\rangle,
\ee
and rewrite Eq.~(\ref{e4}) as the trace in the Fock states:
\begin{gather}
 \begin{split}
\mathrm{Tr}_{\bm\mu}[\exp(\hat x)]&=\sum_{\bm \mu'}\sum_{\bm h'}\langle\bm\mu',\bm h' |\hat {P}_{\bm \mu}\exp(\hat x)|\bm\mu',\bm h'\rangle\\
&=\sum_{\bm n}\langle\bm n |\hat {P}_{\bm \mu}\exp(\hat x)|\bm n\rangle\\
&\equiv\mathrm{Tr}[\hat {P}_{\bm \mu}\exp(\hat x)].
 \end{split}
\end{gather}
Furthermore, we assume that the projection operator $\hat {P}_{\bm \mu}$ can be generated from a exponential quadratic form of bosonic operators, i.e., $\sum_{\bm \mu} t^{\bm \mu}\hat {P}_{\bm \mu}=\hat P(t)$ for $t\in D$ ($D$ is some subset of the complex plane) and
\be
\label{e8}
\hat P(t)=f(t)\exp\left(\bm{\alpha}P(t)\bm{\alpha}^T\right)
\ee
for some $2r\times 2r$ symmetric matrix $P(t)$ and some scalar $f(t)$.  This assumption is satisfied for examples  considered below.
 Then, due to Eqs.~(\ref{e1},  \ref{e8}) the trace formula in the subspace is transformed to the trace of an exponential operator, namely
\begin{gather}
 \begin{split}
\label{e9}
\mathrm{Tr}[\hat {P}(t)\exp(\hat x)]=f(t)\mathrm{Tr}\left[\exp\left(\bm{\alpha}A\bm{\alpha}^T\right)\right],
 \end{split}
\end{gather}
where $A=P(t)+M(x)$. In the Eq.~(\ref{e9}), we have used the property that $\hat {P}_{\bm \mu}, \hat {P}(t)$ commute with $\hat{x}$, since the projection operator is a number in the irreducible subspace.

\section{coherent-state representation}

Coherent-state representation is a well-applied technique to calculate the trace of operators involved bosonic or fermionic operators~\cite{gazeau2009}. It transforms the summation over Fock states to a integral over a complex plane (the latter is convenient to analyze).
In order to simplify Eq.~(\ref{e9}), we introduce  the coherent representation of the bosonic operators~\cite{scully1997,gazeau2009} in the following, i.e., $\bm{z}=(z_1,\cdots,z_n)$,
\begin{gather}
 \begin{split}
\label{e10}
&a_i|\bm z\rangle=z_i|\bm z\rangle,\\
&\langle\bm z|\bm z'\rangle=\prod_{i=1}^{r}\exp\left[-\frac{1}{2}(|z_i|^2+| z'_i|^2-2z_i^*z'_i)\right],\\
&\int\frac{\mathrm d^{2r}\bm{z}}{\pi^r}|\bm z\rangle\langle\bm z|=\sum_{\bm n} |\bm n\rangle\langle\bm n|,
 \end{split}
\end{gather}
where $*$ denotes the complex conjugate, and we have used the abbreviation
\be
\mathrm d^{2r}\bm{z}\equiv\prod_{i=1}^{r}\mathrm d\mathcal{R}(z_i)\mathrm d\mathcal{J}(z_i),
\ee
where $\mathcal{R}(z_i),\mathcal{J}(z_i)$ denote the real part and imaginary part of $z_i$. Since $|\bm z\rangle$ is the eigenstate of the annihilation operators, by using the normal-ordered form of exponential operators in \cite{agrawal1977} and the overcompleteness of the coherent states (Eq.(\ref{e10})), we rewrite Eq.~(\ref{e9}) as a gaussian integral in the coherent representation,
\begin{widetext}
\begin{gather}
 \begin{split}
 \label{e12}
\mathrm{Tr}\left[\exp\left(\bm{\alpha}A\bm{\alpha}^T\right)\right]&=\int \frac{\mathrm d^{2r}\bm{z}}{\pi^r}\langle\bm z|\exp\left(\bm{\alpha}A\bm{\alpha}^T\right)|\bm z\rangle\\
&=[\det(\cosh(\tau A)+\omega\tau\sinh(\tau A))]^{-\frac{1}{2}}\int \frac{\mathrm d^{2r}\bm{z}}{\pi^r}\exp[-\bm{Z}\omega\tau(\coth (\tau A)+\omega\tau)^{-1}\bm{Z}^\dag],
 \end{split}
\end{gather}
\end{widetext}
where $\bm{Z}=(\bm z^*,\bm z)$ and $\bm{Z}^\dag$ is the complex conjugate of $\bm{Z}$.

Now, the trace formula is transformed to a complex gaussian integral (Eq.~(\ref{e12})) which is fully solvable (see Appendix A). Hence  we obtain the result of Eq.~(\ref{e12})
\begin{gather}
 \begin{split}
 \label{e13}
&f(t)\mathrm{Tr}\left[\exp\left(\bm{\alpha}A\bm{\alpha}^T\right)\right]\\
&=f(t)\{(-1)^r \det[\exp(2\tau A)-1]\}^{-\frac{1}{2}}\\&=f(t)\{(-1)^r \det[\exp(2\tau P(t))\exp(2\tau M(x))-1]\}^{-\frac{1}{2}}\\&=\sum_{\bm \mu}t^{\bm \mu}\mathrm{Tr}_{\bm \mu}[\exp(\hat x)]
 \end{split}
\end{gather}
with the convergent condition: the hermitian part of $\omega\tau(\coth (\tau A)+\omega\tau)^{-1}|_{t\in D}$ is positive definite. In the derivation of Eq.~(\ref{e13}), we have used the property $\det[\exp(2\tau A)]=1$ ($\exp(2\tau A)$ is a symplectic matrix). The sign of the square root in Eq.~(\ref{e13}) is determined by two conditions: (1) for $iy\in \mathfrak{h}$, $\hat y$ is hermitian. Its eigenvalues are real. Thus, $\mathrm{Tr}_{\bm \mu}[\exp(\hat y)]>0$; (2) for $x\in\mathfrak{g}_{\mathbb C}$ and $\mathrm{Tr}_{\bm \mu}[\exp(\hat x)]$ converges, $\mathrm{Tr}_{\bm \mu}[\exp(\hat x)]$ is a continuous function of $\hat x$.  Since the determinant is invariant under similar transformation, the trace formula can also be obtained by using the diagonalization of the representation matrices. That is to say for $x,s\in \mathfrak{g}_{\mathbb{C}},y\in\mathfrak{h}_{\mathbb{C}}$, and $x=\exp(-s)y\exp(s)$, if both $\mathrm{Tr}[\hat P(t)\exp(\hat x)], \mathrm{Tr}[\hat P(t)\exp(\hat y)]$ are convergent (see the convergent condition below Eq.~(\ref{e13})), we have
\begin{widetext}
\begin{gather}
 \begin{split}
\mathrm{Tr}[\hat P(t)\exp(\hat x)]&=f(t)\{(-1)^r \det[\exp(2\tau P(t))\exp(2\tau M(x))-1]\}^{-\frac{1}{2}}\\&=f(t)\{(-1)^r \det[\exp(2\tau P(t))\exp(-2\tau M(s))\exp(2\tau M(y))\exp(2\tau M(s))-1]\}^{-\frac{1}{2}}\\&=f(t)\{(-1)^r \det[\exp(2\tau P(t))\exp(2\tau M(y))-1]\}^{-\frac{1}{2}}\\&=\mathrm{Tr}[\hat P(t)\exp(\hat y)],
 \end{split}
\end{gather}
\end{widetext}
and $\mathrm{Tr}_{\bm \mu}[\exp(\hat x)]=\mathrm{Tr}_{\bm \mu}[\exp(\hat y)]=\sum_{\bm h}\exp(y(\bm h))$. In the derivation, we have used the property: $\exp(2\tau M(x))=\exp(-2\tau M(s))\exp(2\tau M(y))\exp(2\tau M(s))$ because $2\tau M$ is the symplectic representation of $\mathfrak{g}_{\mathbb{C}}$~\cite{balian1969}. To illustrate our approach, we calculate the known trace formula of compact groups: the spin system ($\mathrm{SU}(2)$) and the quark system ($\mathrm{SU}(3)$). Besides, as an application, we obtain the result and the convergent condition of the trace formula for $\mathrm{SU}(1,1)$ in Sec.~(\ref{a6}).

\section{Examples}

\subsection{$\mathrm{SU}(2)$}

Let $iJ_j,j=1,2,3$ denote the three generators of $\mathfrak{su}(2)$. We choose $J_3$ as the generator of the Cartan subalgebra. The fundamental representation of $\mathfrak{su}(2)$ (denoted by $\rho$) is given by the Pauli matrices $\sigma_j,j=1,2,3$, i.e., $\rho(J_i)=\sigma_i/2$.

Following Schwinger's construction~\cite{schw2000}, the boson realization of $\mathfrak{su}(2)$ is given by $\hat J_i=\bm \alpha^\dag\rho(J_i)\bm \alpha$, where the dimension of $\bm\alpha$ is 2. Then, the irreducible decomposition of $\mathcal{H}=\{|n_1,n_2\rangle\}$ is $\bigoplus_{2j=0,1,2\cdots}\{|j,m\rangle\}$, where $2j=n_1+n_2,m=-j,-j+1,\cdots,j$, and
\begin{gather}
 \begin{split}
 (\hat{J}_1^2+\hat{J}_2^2+\hat{J}_3^2)|j,m\rangle&=j(j+1)|j,m\rangle\\
\hat{J}_3|j,m\rangle&=m|j,m\rangle.
 \end{split}
\end{gather}

The projection operator onto the subspace $2j$, $\hat P_{2j}=\sum_{m=-j}^{j}|j,m\rangle\langle j,m|$, is generated from a Taylor series
\be
\hat P(t)=\sum_{2j=0}^{\infty}t^{2j}\hat P_{2j}=\exp[(\bm{\alpha}^\dag\cdot\bm{\alpha})\ln t],
\ee
where $t\in[0,\tau]$ for some positive $\tau$ near 0. Then, it follows from Eq.~(\ref{e8}) that $f(t)=t^{-1}$ and
\be
2\tau A=\left(\begin{matrix}
   \ln t+[\rho(x)]^T & 0  \\
   0 & -\ln t-\rho(x)
 \end{matrix}\right).
\ee
For the gaussian integral in Eq.~(\ref{e12}), the converge condition is always satisfied because $\omega\tau(\coth (\tau A)+\omega\tau)^{-1}|_{t\in [0,\tau]}=1/2+O(\tau)$ is positive definite, which is consistent with the fact that $\mathcal{H}_{2j}$ is a finite-dimensional representation of $\mathfrak{su}(2)$.

Finally, from Eq.~(\ref{e13}), we obtain the trace formula of $\mathrm{SU}(2)$:
\begin{gather}
 \begin{split}
\label{e17}
\mathrm{Tr}[\hat P(t)\exp(\hat x)]&=\{\det[e^{-\rho(x)}-t]\}^{-1}\\
&=\sum_{2j=0}^{\infty}t^{2j}\frac{\varepsilon^{2j+1}-\varepsilon^{-2j-1}}{\varepsilon-\varepsilon^{-1}},
 \end{split}
\end{gather}
where $\varepsilon,\varepsilon^{-1}$ are the eigenvalues of $e^{\rho(x)}$. 

\subsection{$\mathrm{SU}(3)$}

Let $iT_j,j=1,2,\cdots,8$ denote the eight generators of $\mathfrak{su}(3)$. We choose $J_3, J_8$ as the generators of the Cartan subalgebra. There are two inequivalent fundamental representations of $\mathfrak{su}(3)$ (denoted by $\rho,\rho'$): (1) $\rho(T_i)=\lambda_i/2$, where $\lambda_j,j=1,2,\cdots,8$ are the Gell-Mann matrices. This is called the ``$3$'' representation; (2) $\rho'(T_i)=-\lambda_i^*/2$. This is called the ``$\bar{3}$'' representation.

According to Ref.~\cite{mathur2001}, the boson realization of $\mathfrak{su}(3)$ is given by
\be
\hat T_i=\bm \alpha^\dag\left(\begin{matrix}
   \rho(T_i) & 0_r  \\
   0_r & \rho'(T_i)
 \end{matrix}\right)\bm \alpha,
\ee
where the dimension of $\bm\alpha$ is 6. Then, the irreducible decomposition of $\mathcal{H}=\{|n_1,\cdots,n_6\rangle\}$ is $\bigoplus_{p,q=0,1,2\cdots}[\mathcal{H}_{(p,0)}\bigotimes\mathcal{H}_{(0,q)}]$, where
\be
p=\sum_{i=1}^{3} n_i,\ \ \ q=\sum_{i=4}^{6} n_i,
\ee
and $(p,q)$ denotes the Dynkin labels of the irreducible representation.

The projection operator onto the subspace $\mathcal{H}_{(p,0)}\bigotimes\mathcal{H}_{(0,q)}$ is
\be
\hat P_{p,q}=\sum_{n_1+n_2+n_3=p,\atop n_4+n_5+n_6=q}|n_1,\cdots,n_6\rangle\langle n_1,\cdots,n_6|.
\ee
It is generated from a Taylor series
\begin{gather}
 \begin{split}
\hat P(t,t')&=\sum_{p,q=0}^{\infty}t^{p}t'^{q}\hat P_{p,q}\\
&=\exp\left[\left(\sum_{i=1}^{3} a^\dag_i a_i\right)\ln t+\left(\sum_{i=4}^{6} a^\dag_i a_i\right)\ln t'\right],
 \end{split}
\end{gather}
where $t,t'\in[0,\tau]$ for some positive $\tau$ near 0. it follows from Eq.~(\ref{e8}) that $f(t)=t^{-1}$ and
\begin{widetext}
\be
2\tau A=\left(\begin{matrix}
   1_3 & 0 & 0 & 0 \\
   0 & 0 & 1_3 & 0 \\
   0 & 1_3 & 0 & 0 \\
   0 & 0 & 0 & 1_3
 \end{matrix}\right)
 \left(\begin{matrix}
   \ln t+[\rho(x)]^T & 0 & 0 & 0  \\
   0 & -\ln t-\rho(x) &0 & 0\\
    0 & 0 & \ln t+[\rho'(x)]^T & 0  \\
    0 & 0 & 0 & -\ln t-\rho'(x)
 \end{matrix}\right)
 \left(\begin{matrix}
   1_3 & 0 & 0 & 0 \\
   0 & 0 & 1_3 & 0 \\
   0 & 1_3 & 0 & 0 \\
   0 & 0 & 0 & 1_3
 \end{matrix}\right).
\ee
\end{widetext}
For the gaussian integral in Eq.~(\ref{e12}), similar to the $\mathrm{SU}(2)$ example,  the convergent condition is always satisfied.

Finally, from Eq.~(\ref{e13}), we obtain the trace formula of $\mathrm{SU}(3)$:
\begin{gather}
 \begin{split}
\label{e24}
\mathrm{Tr}[\hat P(t)\exp(\hat x)]&=\{\det[e^{-\rho(x)}-t]\det[e^{-\rho'(x)}-t']\}^{-1}\\
&=\sum_{p,q=0}^{\infty}t^{p}t'^{q}h_p(\varepsilon_1,\varepsilon_2,\varepsilon_3)h_q(\varepsilon_1^{-1},\varepsilon_2^{-1},\varepsilon_3^{-1}),
 \end{split}
\end{gather}
where $\varepsilon_1,\varepsilon_2,\varepsilon_3$ are the eigenvalues of $e^{\rho(x)}$, and $h_k$ denotes the complete homogeneous symmetric polynomial of degree $k$. Moreover, since~\cite{coleman1964,chat2002}
\be
\mathcal{H}_{(p,0)}\bigotimes\mathcal{H}_{(0,q)}=\bigoplus_{r=0,1,\cdots,\min(p,q)}\mathcal{H}_{(p-r,q-r)},
\ee
it follows from Eq.~(\ref{e24}) that
\begin{gather}
 \begin{split}
 \label{e25}
&\mathrm{Tr}_{(p,q)}[\exp(\hat x)]\\
&=\mathrm{Tr}_{(p,0)\bigotimes(0,q)}[\exp(\hat x)]-\mathrm{Tr}_{(p-1,0)\bigotimes(0,q-1)}[\exp(\hat x)]\\
&=h_p(\varepsilon_1,\varepsilon_2,\varepsilon_3)h_q(\varepsilon_1^{-1},\varepsilon_2^{-1},\varepsilon_3^{-1})-\\
&\ \ \ \ h_{p-1}(\varepsilon_1,\varepsilon_2,\varepsilon_3)h_{q-1}(\varepsilon_1^{-1},\varepsilon_2^{-1},\varepsilon_3^{-1}).
 \end{split}
\end{gather}
In comparison with the Weyl character formula~\cite{weyl1953,fulton2004}, Eq.~(\ref{e25}) indicates some identity relations between the complete homogeneous symmetric polynomials.

\section{Application: trace formula of $\mathrm{SU}(1,1)$}
\label{a6}

$\mathrm{SU}(1,1)$ is the most elementary  non-compact group~\cite{perelomov1986}. Its irreducible unitary representation is infinite dimensional with equidistant basis vectors, which is just the energy level of quantum harmonic oscillators.
Besides, the quantum many-body systems with $\mathrm{SU}(1,1)$ as the dynamical group is of particular interest~\cite{beau2016,deng2018,myers2021} since it is a broad family of interacting many-body systems that are analytically solvable. As an example of this family, the Calogero–Sutherland model clearly shows the transformation of statistical behaviors between bosons and fermions through changing the strength of interaction, which is known as the fractional exclusion statistics~\cite{murthy1994}. Other examples include the non-interacting bosons or fermions, the Tonks-Girardeau gas, two-dimensional Bose-Einstein condensates with s-wave contact interactions~\cite{pita1997}, and the unitary Fermi gas~\cite{werner2006}, which have been realized in ultracold atom laboratories~\cite{kino2004,pare2004,hara2002,saint2019,deng2018}.

Let $iK_j,j=1,2,3$ denote the three generators of $\mathfrak{su}(1,1)$. We choose $K_3$ as the generators of the Cartan subalgebra. The fundamental representation of $\mathfrak{su}(1,1)$ (denoted by $\rho$) is given by the Pauli matrices: $\rho(K_1)=i\sigma_2/2,\rho(K_2)=-i\sigma_1/2,\rho(K_3)=\sigma_3/2$.

According to Ref.~\cite{gerry1991}, the two-mode realization of $\mathfrak{su}(2)$ is given by
\begin{gather}
 \begin{split}
\hat K_1&=\frac{a^\dag_1a^\dag_2+a_1a_2}{2}\\
\hat K_2&=\frac{a^\dag_1a^\dag_2-a_1a_2}{2i}\\
\hat K_3&=\frac{a^\dag_1a_1+a_2^\dag a_2+1}{2}\\
 \end{split}
\end{gather}
 Then, the irreducible decomposition of $\mathcal{H}=\{|n_1,n_2\rangle\}$ is
 \be
 \mathcal{H}=\{|k=1/2,m,s=0\rangle\}\bigoplus_{2k=1,2,\cdots\atop s=\pm}\{|k,m,s\rangle\},
 \ee
 where  $2k=|n_1-n_2|+1,m=0,1,\cdots$. And the following relations are satisfied
\begin{gather}
 \begin{split}
 \label{e30}
 (\hat{K}_1^2-\hat{K}_2^2-\hat{K}_3^2)|k,m,s\rangle&=k(k-1)|k,m,s\rangle\\
\hat{K}_3|k,m,s\rangle&=(k+m)|k,m,s\rangle\\
\hat{K}_+|k,m,s\rangle&=\sqrt{(m+1)(m+2k)}|k,m+1,s\rangle\\
\hat{K}_-|k,m,s\rangle&=\sqrt{m(m+2k-1)}|k,m-1,s\rangle,
 \end{split}
\end{gather}
where $\hat{K}_\pm=\hat K_1\pm i\hat K_2$.
Here, we set $n_1>n_2$ for $|k,m,+\rangle$ and $n_1<n_2$ for $|k,m,-\rangle$.

The projection operator onto the subspace $\mathcal{H}_{k,s}$ is
\be
\hat P_{k,s}=\sum_{m=0}^{\infty}|k,m,s\rangle\langle k,m,s|.
\ee
It is generated from a Laurent series (different from the Taylor series in the compact groups $\mathrm{SU}(2)$ and $\mathrm{SU}(3)$)
\begin{gather}
 \begin{split}
  \label{e301}
\hat P(t)=&t^{-1}\hat P_{1/2,0}+\sum_{2k=1}^{\infty}t^{-2k-1}\hat P_{k,-}+\sum_{2k=1}^{\infty}t^{2k-1}\hat P_{k,+}\\
=&\exp[(a_1^\dag a_1-a_2^\dag a_2)\ln t],
 \end{split}
\end{gather}
where the domain of $t$ is shown below. Then, it follows from Eq.~(\ref{e8}) that $f(t)=1$ and
\be
2\tau A=\left(\begin{matrix}
   1 & 0 & 0 & 0 \\
   0 & 0 & 1 & 0 \\
   0 & 0 & 0 & 1 \\
   0 & 1 & 0 & 0
 \end{matrix}\right)
 \left(\begin{matrix}
   \ln t+\rho(x)  & 0  \\
    0  & -\ln t+\rho(x)
 \end{matrix}\right)
 \left(\begin{matrix}
   1 & 0 & 0 & 0 \\
   0 & 0 & 0 & 1 \\
   0 & 1 & 0 & 0 \\
   0 & 0 & 1 & 0
 \end{matrix}\right).
\ee

Finally, from Eq.~(\ref{e13}), we obtain the trace formula of $\mathrm{SU}(1,1)$:
\be
\label{e34}
\mathrm{Tr}[\hat P(t)\exp(\hat x)]=\frac{-t}{\det[e^{\rho(x)}-t]}=\frac{-t}{(t-\varepsilon)(t-\varepsilon^{-1})},
\ee
where $\varepsilon,\varepsilon^{-1}$ are the eigenvalues of $e^{\rho(x)}$. For $|\varepsilon|=1$, the Laurent series of Eq.~(\ref{e34}) does not exist and our approach is invalid. For $|\varepsilon|\neq 1$, without loss of generality, setting $|\varepsilon|<1$, the Laurent series of Eq.~(\ref{e34}) reads
\be
\frac{-t}{(t-\varepsilon)(t-\varepsilon^{-1})}=\sum_{n=-\infty}^{\infty}t^{n}\frac{\varepsilon^{n+1}}{1-\varepsilon^2}
\ee
which is valid for $|\varepsilon|<|t|<|\varepsilon|^{-1}$. Hence, the convergent condition of Eq.~(\ref{e34}) reads: the hermitian part of $\omega\tau(\coth (\tau A)+\omega\tau)^{-1}|_{|\varepsilon|<|t|<|\varepsilon|^{-1}}$ is positive definite. According to Eq.~(\ref{e301}), the trace formula of $\mathrm{SU}(1,1)$ is
\be
\mathrm{Tr}_{2k}[\exp(\hat x)]=\frac{\varepsilon^{2k}}{1-\varepsilon^2},
\ee
which is also generalized to the case that $2k$ is a positive real number by using the Barut-Girardello coherent states (see Appendix B).

\section{Conclusion}

Based on the boson realization of (complex) Lie groups, we propose a unified approach to the calculation of the trace of exponential operators. Besides the known trace formula of compact groups (Weyl character formula), we also obtain the result and the convergent condition of the trace for non-compact groups by using this approach. A useful conclusion is that under convergent conditions, the trace of exponential operators in the Hilbert space is equal to a determinant involving the lower-dimensional representation of the group element. Since two  elementary operators in quantum mechanics (the time evolution operator and the density operator of equilibrium state) are exponential operators, we hope our approach will be widely applied in the future in the study of equilibrium thermodynamics and nonequilibrium process of quantum (many-body) systems.

\begin{acknowledgments}
We thank Jinpeng An for helpful discussions.
This work is supported by the National Natural Science Foundation
of China (NSFC) (Grants No. 12088101), NSAF (No. U1930402, and No. U1930403). Z. Y. Fei acknowledges support from China Postdoctoral Science Foundation (Grant No. 2021M700359).

\end{acknowledgments}

 \renewcommand{\theequation}{A.\arabic{equation}}

 \setcounter{equation}{0}

 \section*{Appendix A: evaluation of the complex gaussian integral}
Let
\be
\label{a1}
\Lambda(B)=\int \frac{\mathrm d^{2r}\bm{z}}{\pi^r}\exp\left(-\frac{1}{2}\bm{Z}B\bm{Z}^{\dag}\right),
\ee
where $\omega B$ is a $2r\times 2r$ complex symmetric matrix.

Every $B$ has a unique decomposition $B = B'
+iB''$, with hermitian $B'$ and $B''$, and we call $B'$
the hermitian part of $B$. Then, for invertible $B'$, the integral in Eq.~(\ref{a1}) converges if and only if $B'$ is positive definite and the result reads
\be
\Lambda(B)=[\det(B)]^{-\frac{1}{2}}.
\ee
\emph{Proof}. Let $z_k=x_{2k-1}+ix_{2k},\bm X=(x_1,\cdots,x_{2r})$. Then, Eq.~(\ref{a1}) is rewritten as
\begin{gather}
 \begin{split}
\label{a3}
&\Lambda(B)=\pi^{-r}\int \prod_{i=1}^{2r}\mathrm dx_i\exp\left(-\bm{X}UBU^{\dag}\bm{X}^{T}\right)\\
&=\pi^{-r}\int \prod_{i=1}^{2r}\mathrm dx_i\exp\left(-\bm{X}UB'U^{\dag}\bm{X}^{T}-i\bm{X}UB''U^{\dag}\bm{X}^{T}\right),
 \end{split}
\end{gather}
where $\bm Z=\sqrt{2}\bm{X}U$, and both $UB'U^{\dag},UB''U^{\dag}$ are real symmetric matrices (notice that $U^{\dag}=U^{-1}=\omega U^T$).

We proceed in three steps: (1) If $B=1$, then $\Lambda(1)=\det(1)=1$; (2)  Eq.~(\ref{a3}) is convergent if and only if it is absolutely convergent~\cite{beesack1970}, i.e., $B'$ is positive definite; (3) If $B''=0$, $B'$ is positive definite, $UB'U^{\dag}$ can be factorized as $UB'U^{\dag}=LL^T$, where $L$ is a real lower triangular matrix with positive diagonal entries (the Cholesky decomposition~\cite{horn1985}). Setting $\bm X'=(x'_1,\cdots,x'_r)=\bm XL$, we find that the Jacobian of the transformation is $[\det(L)]^{-1}$.
Hence, $\Lambda(B)=[\det(L)]^{-1}\Lambda(1)=[\det(B)]^{-\frac{1}{2}}$. This result is still valid when $B''\neq 0$ by following the similar procedure of analytic continuation in the appendix of Ref.~\cite{bargmann1962}.

 \renewcommand{\theequation}{B.\arabic{equation}}

 \setcounter{equation}{0}

 \section*{Appendix B: the trace formula of $\mathrm{SU}(1,1)$ using the Barut-Girardello coherent states}

For $k$ is a positive real number, let the states $|k,m\rangle$ which satisfy the relations in Eq.~(\ref{e30}) denote the (projective) representation of $\mathrm{SU}(1,1)$. Then, we define eigenstates $|k,z\rangle$ of the lowering operator $\hat K_-$ as:
\be
\hat K_-|k,z\rangle=z|k,z\rangle,
\ee
where $z$ is an arbitrary complex number. Thus, $\{|k,z\rangle\}$ are called the Barut-Girardello states~\cite{barut1971} and can be decomposed over the orthonormal state basis $\{|k,m\rangle\}$,
\be
\label{b2}
|k,z\rangle=\frac{z^{k-1/2}}{\sqrt{I_{2k-1}(2|z|)}}\sum_{m=0}^{\infty}\frac{z^m}{m!\Gamma(m+2k)}|k,m\rangle
\ee
where $\Gamma$ denotes the Euler gamma function and  $I_{2k-1}$ denotes the modified Bessel function of the first kind. The Barut-Girardello coherent states are overcomplete, i.e.,
\begin{gather}
 \begin{split}
&\langle k,z|k,z'\rangle=\frac{I_{2k-1}(2\sqrt{z^*z'})}{\sqrt{I_{2k-1}(2|z|)I_{2k-1}(2|z'|)}},\\
&\int\mathrm{d}\mu(k,z) |k,z\rangle\langle k,z|=\sum_{m=0}^{\infty} |k,m\rangle\langle k,m|
 \end{split}
\end{gather}
with the measure
\begin{gather}
 \begin{split}
\mathrm d\mu(k,z)&=\frac{2}{\pi}K_{2k-1}(2|z|)I_{2k-1}(2|z|)\mathrm{d}^2 z,\\
\mathrm{d}^2 z&=\mathrm d\mathcal{R}(z)\mathrm d\mathcal{J}(z).
 \end{split}
\end{gather}
Here,  $K_{2k-1}$ denotes the modified Bessel function of the second kind.

In general, the exponential operator $\exp(x)$ can be rewritten in the following form:
\be
e^x=e^{\lambda_+K_+}e^{\lambda_3 K_3}e^{\lambda_-K_-},
\ee
where $\lambda_3,\lambda_\pm$ are obtained by using the composition formula of $\mathrm{SU}(1,1)$ in Ref.~\cite{martinez2020}. Then, the trace formula for $\mathrm{SU}(1, 1)$ reads
\begin{gather}
 \begin{split}
 \label{b6}
&\mathrm{Tr}_{2k}[e^{\lambda_+K_+}e^{\lambda_3 K_3}e^{\lambda_-K_-}]\\
&\equiv  \sum_{m=0}^{\infty} \langle k,m|e^{\lambda_+K_+}e^{\lambda_3 K_3}e^{\lambda_-K_-}|k,m\rangle\\
                                                              &=\int\mathrm{d}\mu(k,z) \langle k,z|e^{\lambda_+K_+}e^{\lambda_3 K_3}e^{\lambda_-K_-}|k,z\rangle\\
                                                              &=\int\mathrm{d}\mu(k,z) \langle k,z|e^{\lambda_3 K_3}|k,z\rangle e^{\lambda_+z^*+\lambda_-z}\\
                                                              &=\int\mathrm{d}\mu(k,z)  \frac{I_{2k-1}(2e^{\lambda_3/2}|z|)}{I_{2k-1}(2|z|)}e^{\lambda_3/2+\lambda_+z^*+\lambda_-z}
 \end{split}
\end{gather}
where we have used Eq.~(\ref{b2}) and the series expansion of $I_{\nu}(2x)$:
\be
I_{\nu}(2x)=\sum_{n=0}^{\infty}\frac{x^{2n+\nu}}{n!\Gamma(n+\nu+1)}
\ee
in the derivation.

Using the polar coordinate of the complex number $z=re^{i\phi}, r\in[0,\infty), \phi\in[0,2\pi)$ and first integrate Eq.~(\ref{b6}) over $\phi$, we finally obtain the integral of the product of three Bessel functions:
\begin{gather}
 \begin{split}
&\mathrm{Tr}_{2k}[e^{\lambda_+K_+}e^{\lambda_0 K_0}e^{\lambda_-K_-}]\\
&=4e^{\lambda_0/2}\int_{0}^{\infty}\mathrm dr r I_0(2\sqrt{\lambda_+\lambda_-}r)I_{2k-1}(2e^{\lambda_0/2}r)K_{2k-1}(2r).
 \end{split}
\end{gather}
This integral converges when each of the four $\mathrm{Re}(2\pm 2\sqrt{\lambda_+\lambda_-}\pm2e^{\lambda_0/2})$ is positive~\cite{bailey1936}. For the convergent case, using Eq.~(3.1) in Ref.~\cite{bailey1936} and Eq.~(44a) in Ref.~\cite{Schlosser2013}, we obtain the result of the integral
\be
\mathrm{Tr}_{2k}[e^{\lambda_+K_+}e^{\lambda_3 K_3}e^{\lambda_-K_-}]=\frac{\varepsilon^{2k}}{1-\varepsilon^2},
\ee
where $\varepsilon$ satisfies
\be
\label{b10}
\varepsilon^2+e^{-\lambda_0/2}(\lambda_+\lambda_--e^{\lambda_0}-1)\varepsilon+1=0.
\ee
Connecting Eq.~(\ref{b10}) with the fundamental representation matrix $e^{\lambda_+\rho(K_+)}e^{\lambda_3 \rho(K_3)}e^{\lambda_-\rho(K_-)}=e^{\rho(x)}$, we find that Eq.~(\ref{b10}) is just the characteristic polynomial of $e^{\rho(x)}$. Hence we have
\be
\mathrm{Tr}_{2k}[e^{x}]=\frac{\varepsilon^{2k}}{1-\varepsilon^2},
\ee
where $\varepsilon$ is one of the eigenvalues of $e^{\rho(x)}$.


\begin{thebibliography}{99}

\bibitem{yang1952} C. N. Yang and T. D. Lee, Phys. Rev. 87, 404 (1952).
\bibitem{lee1952} T. D. Lee and C. N. Yang,  Phys. Rev. 87, 410 (1952).
\bibitem{fisher1965} M. E. Fisher, The nature of critical points, in \emph{Lectures in Theoretical Physics. Vol.
VIIC | Statistical Physics, Weak Interactions, Field Theory}, ed. W. E. Brittin (Univ. Colorado Press, Boulder, 1965).

\bibitem{aq2000} J. Kurchan, arXiv preprint cond-mat/0007360 (2000).
\bibitem{ja2000} H. Tasaki, arXiv preprint cond-mat/0009244 (2000).
\bibitem{kinder2003} M. Kindermann, Y. V. Nazarov, Y. V.  Full counting statistics in electric circuits. In \emph{Quantum Noise in Mesoscopic Physics} (pp. 403-427). (Springer, Dordrecht 2003).
\bibitem{goussev2012} A. Goussev, R. A. Jalabert, H. M. Pastawski, D. Wisniacki,  arXiv preprint arXiv:1206.6348 (2012).

\bibitem{peng2015} X. Peng, H. Zhou, B. B. Wei, J. Cui, J. Du, and R. B. Liu,  Phys. Rev. Lett. \textbf{114}, 010601 (2015).
\bibitem{mazzola2013} L. Mazzola, G. De Chiara, and M. Paternostro, Phys. Rev. Lett. \textbf{110}, 230602 (2013).

\bibitem{levitov1996} L. S. Levitov and H. Lee, Journal of Mathematical Physics 37, 4845 (1996).
\bibitem{klick2003} I. Klich, An elementary derivation of Levitov´s formula. In \emph{Quantum Noise in Mesoscopic Physics} (pp. 397-402) (Springer, Dordrecht 2003).
\bibitem{fei2019} Z. Y. Fei and H. T. Quan, Phys. Rev. Research \textbf{1}, 033175 (2019).

\bibitem{weyl1953} H. Weyl, \emph{The Classical Groups, 2nd ed.},  p. 320. (Princeton Univ. Press, Princeton,  1953)
\bibitem{fulton2004} W. Fulton, J. Harris, \emph{Representation Theory: A First Course}, p. 551., (Springer, New York,  2004).
\bibitem{fall2015} M. Fallbacher, Nuclear Physics B \textbf{898}, 229–247 (2015).
\bibitem{knapp2016} W. A. Knapp, \emph{Representation theory of semisimple groups.} (Princeton university press, 2016).
\bibitem{kirillov2012} A. A. Kirillov,  \emph{Elements of the Theory of Representations.} ( Springer Science \& Business Media, 2012).

\bibitem{gerry1991} C. C. Gerry, Journal of the Optical Society of America B \textbf{8}(3), pp. 685-690 (1991).
\bibitem{mathur2010} M. Mathur, I. Raychowdhury, and R. Anishetty, J. Math. Phys. \textbf{51}, 093504 (2010).
\bibitem{beau2020} M. Beau, and A. del Campo, Entropy \textbf{2020}, 22, 515 (2020).

\bibitem{balian1969} R. Balian and E. Brezin, II Nuovo Cimento B 64, 37 (1969).
\bibitem{hofmann1978} K. H. Hofmann, and A. Mukherjea, Math. Ann. \textbf{234}, 263-273 (1978).
\bibitem{gazeau2009} J.-P. Gazeau, \emph{Coherent States in Quantum Physics}, (2009 WILEY-VCH Verlag GmbH \&
Co. KGaA, Weinheim)

\bibitem{scully1997} M. O. Scully, M. S. Zubairy, \emph{Quantum optics}. (Cambridge University Press, 1997).
\bibitem{agrawal1977} G. P. Agrawal and C. L. Mehta, J. Math. Phys. \textbf{18}, 408 (1977).
\bibitem{schw2000} \emph{A Quantum Legacy—Seminal Papers of Julian Schwinger}, edited by K. A. Milton, pp. 173 (World Scientific, Singapore, 2000).
\bibitem{mathur2001} M. Mathur, D. Sen, J. Math. Phys. \textbf{42}, 4181 (2001).
\bibitem{coleman1964} S. Coleman, J. Math. Phys. \textbf{5}, 1343-1964 (1964).
\bibitem{chat2002} S. Chaturvedi and N. Mukunda, J. Math. Phys. \textbf{43}, 5278 (2002).
\bibitem{perelomov1986} A. Perelomov, \emph{Generalized Coherent States and Their Applications.}, (Springer-Verlag Berlin Heidelberg, 1986).



\bibitem{beau2016} M. Beau, J. Jaramillo, and A. del Campo, Entropy \textbf{18}, 168 (2016).
\bibitem{deng2018} S. Deng, et al., Sci. Adv. \textbf{4} eaar5909 (2018).
\bibitem{myers2021} N. M. Myers, J. McCready, and S. Deffner, Symmetry  \textbf{13}, 978 (2021).
\bibitem{murthy1994} M. V. N. Murthy, and R. Shankar, Phys. Rev. Lett. \textbf{73}(25), 3331 (1994).

\bibitem{pita1997} L. P. Pitaevskii and A. Rosch, Phys. Rev. A \textbf{55}, R853(R) (1997).
\bibitem{werner2006} F. Werner, and Y. Castin, Phys. Rev. A \textbf{74}, 053604 (2006).
\bibitem{kino2004} T. Kinoshita, T. Wenger, D. S. Weiss, Science \textbf{305}(5687), 1125–1128 (2004).
\bibitem{pare2004} B. Paredes, et al., Nature \textbf{429}, 277–281 (2004).
\bibitem{saint2019} R. Saint-Jalm, et. al., Phys. Rev. X \textbf{9}, 021035 (2019).
\bibitem{hara2002} K. M. O'Hara, et. al, Science \textbf{298}, 2179 (2002).
\bibitem{beesack1970} P. R. Beesack, Mathematics Magazine, \textbf{43}(3), pp. 113-123 (1970).
\bibitem{horn1985} R. A. Horn, C. R. Johnson,   \emph{Matrix Analysis}. (Cambridge University Press, 1985).
\bibitem{bargmann1962} V. Bargmann, Rev. Mod. Phys, \textbf{34}(4), 829 (1962).

\bibitem{barut1971} A. O. Barut, and L. Girardello,  Commun. Math. Phys. \textbf{21}, 41–55 (1971).
\bibitem{martinez2020} D. M. Tibaduiza, A. H. Arag\~{a}o, C. Farina, C. A. D. Zarro, Phys. Lett. A \textbf{384}, 126937 (2020).
\bibitem{bailey1936} W. N. Bailey, Journal of the London Mathematical Society, Volume s1-11, Issue 1, January 1936, Pages 16–20.
\bibitem{Schlosser2013} M. J. Schlosser, \emph{Computer Algebra in Quantum Field Theory}, (Springer-Verlag Wien, 2013), pp. 305–324.

%
%
%
%
%
%
%
%
%
%
%
%
%
%
%
%
%
%
%
%
%
%
%
%
%
%


 \end{thebibliography}
\end{document}